\documentclass[usenatbib,usegraphicx]{mn2e}
\usepackage{times}

\usepackage{multicol}
\usepackage{lscape}
\usepackage{longtable,lscape}
\usepackage{multirow}
\usepackage{color}
\usepackage{cancel}

\newcommand{\dd}[1]{{\mbox{\scriptsize #1}}}
\newcommand{\ccol}[1]{\multicolumn{1}{c}{#1}}

\newcommand{\us}{\mbox{USNO-B}}
\newcommand{\arcdeg}{$^{\circ}$}

\title[Deep blank field catalogue]{Deep blank field catalogue for medium- and large-size telescopes}

\author[F.M.~Jim\'{e}nez-Esteban et al.]{F.M.~Jim\'{e}nez-Esteban,$^{1,2,3}$\thanks{E-mail: fran.jimenez-esteban@cab.inta-csic.es} A.~Cabrera-Lavers,$^{4,5}$ N. Cardiel,$^{6}$ and J.M.~Alacid$^{1,2}$\\
$^{1}$Centro de Astrobiolog\'{\i}a (INTA-CSIC), Departamento de 
Astrof\'{\i}sica, PO Box 78, E-28691, Villanueva de la Ca\~nada, Madrid, Spain\\
$^{2}$Spanish Virtual Observatory, Spain\\
$^{3}$Saint Louis University, Madrid Campus, Division of Science and Engineering, Avenida~del~Valle 34, E-28003 Madrid, Spain \\
$^{4}$Instituto de Astrof\'{\i}sica de Canarias, E-38205 La Laguna, Tenerife, Spain \\
$^{5}$GTC Project, E-38205 La Laguna, Tenerife, Spain \\
$^{6}$Departamento de Astrof\'{\i}sica y CC.\ de la Atm\'{o}sfera, Facultad de Ciencias F\'{\i}sicas, Avenida Complutense s/n, E-28040  Madrid, Spain}

\begin{document}

\date{Accepted \ldots Received \ldots; in original form \ldots}

\pagerange{\pageref{firstpage}--\pageref{lastpage}} \pubyear{2012}

\maketitle

\label{firstpage}

\begin{abstract}
The observation of blank fields, defined as regions of the sky that
are devoid of stars down to a given threshold magnitude, constitutes
one of the most relevant calibration procedures required for the
proper reduction of astronomical data obtained following typical
observing strategies. In this work, we have used the Delaunay
triangulation to search for deep blank fields throughout the whole
sky, with a minimum size of 10\arcmin\ in diameter and an increasing
threshold magnitude from 15 to 18 in the {\it R} band of the USNO-B
Catalog of the United States Naval Observatory. The result is a
catalogue with the deepest blank fields known so far. A short sample
of these regions has been tested with the 10.4m Gran Telescopio
Canarias, and it has been shown to be extremely useful for medium and
large size telescopes. Because some of the regions found could also be
suitable for new extragalactic studies, we have estimated the galactic
extinction in the direction of each deep blank field. This catalogue
is accessible through the Virtual Observatory tool TESELA, and the
user can retrieve - and visualize using Aladin - the deep blank fields
available near a given position in the sky.
\end{abstract}


\section{Introduction}
 \label{section:introduction}

In observational astrophysics, and in particular focusing in imaging
acquisition through the optical window, the data treatment is
fundamental to minimize the influence of data acquisition
imperfections on the estimation of the desired astronomical
measurements (e.g.\ \citealt{Gilliland92}). For this purpose, image
flat-fielding and sky subtraction constitute two of the most common
and important reduction steps. An inadequate flat-fielding or sky
subtraction easily leads to the introduction of systematic
uncertainties in the data. Therefore, the observation of blank fields
(BFs), defined as regions of the sky that are devoid of stars down to
a given threshold magnitude ($m_\dd{th}$), is a very important aspect
in astronomical observations. In our previous paper \cite{Cardiel11}
(hereafter Paper~I), we have already discussed in detail the relevance
of BFs in astronomical observations, and we refer the interested
reader to the introduction of that paper.

In Paper~I, we presented the first systematic all-sky catalogue of BFs
so far. The method used to identify the BFs was based on the Delaunay
triangulation on the surface of a sphere, using stars brighter than 11
mag taken from the \mbox{Tycho-2} \citep{Hog00} catalogue as the nodes
for triangulation. In addition, a new Virtual Observatory (VO) tool
named TESELA\footnote{\tt
  http://sdc.cab.inta-csic.es/tesela/index.jsp}, accessible through
the internet, was created so that users can retrieve, and visualize,
the BFs available near a given position in the sky.

Another commonly used resource is the collection by Marco
Azzaro\footnote{\tt http://www.ing.iac.es/$\sim$meteodat/blanks.htm},
consisting of a short list of 38 BFs voids of stars up to 10--16
mag. Although, in general, deeper than the BFs of Paper~I, these are
very few, and most of them are in the Northern hemisphere; there are
only a few with negative declination, and none bellow
$-20^{\circ}$. So, this BF collection is not well suited for
telescopes located in the Southern hemisphere.

Nevertheless, even the Azzaro BFs have been shown to be shallow for
large-class telescopes. For example, in the case of the Gran
Telescopio Canarias\footnote{\tt http://www.gtc.iac.es} (GTC), a 10-m
class telescope, equipped with the Optical System for Imaging and Low
Resolution Integrated Spectroscopy (OSIRIS\footnote{\tt
  http://www.gtc.iac.es/en/pages/instrumentation/osiris.php},
\citealp{Cepa00}), it is usually necessary to have access to BFs free
of stars down to 17.5 mag. This requirement will be even more
demanding for the future generation of telescopes, such as European
Extremely Large Telescope\footnote{\tt 
  http://www.eso.org/public/teles-instr/e-elt.html} (E-ELT).

One method of identifying low-populated fields is to use a binned-up
smoothed map of the distribution of the stars in the sky. In this
method, it is necessary to divide the sky in small patches and to
identify those with few, or even no stars, down to a threshold
magnitude. However, with this method, a BF would not be identified
unless the patch was well centered on the region free of stars. In
addition, the size of the BF would not be the maximum possible because
it is arbitrarily defined. Thus, the procedure should be repeated many
times with both a different centring scheme and a different patch
radius; this would be very demanding with respect to CPU time. These
two drawbacks can be overcome with the sky tessellation method we have
developed, which has been extensively described in Paper~I.

Thus, in this work we have applied Delaunay triangulation to the
USNO-B Catalog of the United States Naval Observatory, we have
generated acatalogue of deep blank fields (DBFs) to varying limiting
magnitudes from 15 to 18 in steps of 1 mag. In order to validate our
method, we have tested some DBFs found at the GTC and we have compared
these with the BFs in Azzaro catalogue. Finally, in order to
facilitate the use of this new catalogue, we have incorporated it into
TESELA, which now provides a simple interface that allows the user to
retrieve the list of either the BFs from \mbox{Tycho-2} or the DBFs
from \us available near a given position in the sky.

In Section~\ref{section:catalogues}, we describe the method used to
create the catalogues, the defects found in the \us\ Catalog and the
results of the tessellation. Section~\ref{section:GTC}, we describe
the use of the DBFs catalogue on the GTC. And, finally, in
Section~\ref{section:TESELA}, we describe the implementation of
TESELA.

\section{The catalogue}
  \label{section:catalogues}

\subsection{Tessellating the sky}
  \label{section:tessellating}

The Delaunay triangulation \citep{Delaunay34} consists of the
subdivision of a geometric object (e.g.\ a surface or a volume) into a
set of simplices. In particular, for the Euclidean planar
(two-dimensional) case, given a set of points, also called nodes,
Delaunay triangulation becomes a subdivision of the plane into
triangles, whose vertexes are nodes. For each of these triangles, it
is possible to determine its associated circumcircle, which is the
circle passing exactly through the three vertices of the
triangle. Interestingly, in a Delaunay triangulation, all the
triangles satisfy the empty circumcircle interior property, which
means that all the circumcircles are empty (i.e. there are no nodes
inside any of the computed circumcircles). See Paper~I for an extended
description of the method.

Delaunay triangulation can be applied to the two-dimensional surface
of a three-dimensional celestial sphere, using as nodes the locations of
the stars down to a given $m_\dd{th}$. Then, the empty circumcircle
interior property of the Delaunay triangulation provides a
straightforward method for a systematic search of regions void of
stars brighter than a certain magnitude.

In this work we applied the Delaunay triangulation to the sources of
the \us\ Catalog \citep{Monet03}. The \us\ Catalog contains
astrometric and photometric information for more than a billion stars
and galaxies in the sky. The data were taken from various sky surveys
over 50 years, and it is expected to be complete up to magnitude {\it
  V}~=~21~mag. Typical uncertainties are 200~mas for the positions at
J2000 and 0.3~mag in photometry. Photometric data consists of {\it V,
  R}, and {\it I} optical passbands. We used the {\it R} band (between
600 and 750 nm) as the magnitude reference. For most of the sources,
the \us\ Catalog provides the {\it R} magnitude ($m_\dd{R}$) at two
different epochs. To assign a single $m_\dd{R}$ to these objects,
either we have averaged out both $m_\dd{R}$ when their values are
similar ($\Delta m_\dd{R}$\,$<$1) or selected the lower value (i.e.,
the brightest value) when there is a larger difference.

In principle, it is straightforward to obtain a list of BFs for the
whole celestial sphere by using as input a stellar catalogue including
stars down to a given magnitude. However, as explained in Paper~I,
because of the exponential growth of the number of objects with
increasing limiting magnitude, tessellating the whole sky with
$m_\dd{th}$ above 11 mag requires too many computer resources
(in form of memory and computing time), even with the approach of
subdividing the sky into many smaller subregions.

Therefore, to create the DBFs catalogue, we have followed a
methodology that is different from the one adopted in Paper~I. Instead
of tessellating the whole sky at once, we used the following
workflow. (i) We selected all BFs with a size larger than 10\arcmin,
using the deepest catalogue available at that moment. (ii) We
individually tessellated each of these regions, as in Paper~I, using
the positions of the \us\ objects down to a $m_\dd{th}$ as the nodes
of the triangulation. (iii) To create the new DBFs catalogue, from the
new collection of deeper BFs we selected those with a size larger than
10\arcmin. (iv) We repeated the whole process, increasing $m_\dd{th}$
from 15 to 18 in steps of 1 mag.

The initial collection of BFs was obtained from Paper~I, consisting of
BFs void of stars down to 11 mag in both \mbox{Tycho-2}
passbands. This provided \mbox{$1\,477\,348$} initial regions with a
size larger than 10\arcmin. This size was selected to ensure that the
regions were always larger than the field of view of the OSIRIS
instrument used at the GTC (see Sect.\ref{section:GTC}). Tessellating
this huge number of initial regions requires months of execution
time. In order to obtain the results as soon as possible so they could
be used in the nightly operation of the GTC, we proceeded with the
Northern (Dec\,$>$\,$-$30\arcdeg) sky in strips 1 h wide in RA.  When
the tessellation of the Northern sky was finished, we tessellated the
Southern (Dec\,$\le$\,$-$30\arcdeg) sky in the same way. Finally we
merged all DBFs found in one catalogue.

It was very common that the tessellation process identified two
overlaping DBFs which centers were separated by less than 0.3 times
the smaller DBF radius. Although both DBFs are correct, the two refer
to almost the same region of the sky. Thus, to facilitate the use of
the catalogue, we removed the DBF with the smaller radius from the
catalogue, retaining the one with the larger size.

A final evaluation of each DBF of the catalogue was made. We counted
the total number of stars inside the DBF with both $m_\dd{R}$ lower
than 17.5 mag (a requirement of the GTC; see
Section\,\ref{section:GTC}) and lower than 22.0 mag. Then, we
estimated the integrated {\it R} flux ($F_\dd{int}$) as the sum of the
individual fluxes of all star in the DBF with
$m_\dd{R}$\,$<$\,22.0\,mag. These quantities provided a measurement of
the depth of the DBF.

To verify the validity of the method, and taking advantage of the
Aladin\footnote{\tt http://aladin.u-strasbg.fr/} software
\cite[][]{Bonnarel00}, we visually checked all the 79 DBFs of 18
$m_\dd{th}$ found previously in order to give them to the operation
staff of the GTC. This VO tool allows users to visualize and analyse
digitized astronomical images, and superimpose entries from
astronomical catalogues or data bases available from the VO
services. Using Aladin, we displayed the DSS-red images of the DBF
region and we superimposed the \us\ data of the region. This exercise
was very useful in the detection of unknown defects in the
\us\ Catalog (see Sect.\ref{section:defects}). Of the 79 DBFs, 5
corresponded to a dark cloud, so they were removed from the final
catalogue. Similarly, we removed any DBFs corresponding to the same
dark cloud of any $m_\dd{th}$.


\subsection{Defects in the USNO-B Catalogue}
  \label{section:defects}

The \us\ Catalog was constructed from photographic plates, taken from
several different surveys, which were uniformly scanned, and whose
sources were extracted in an automatic way from the scanned data. The
original plate images contained many defects and artifacts that
logically affected the final \us\ data. That said, so far there is no
other full sky catalogue deep enough for our project. So, the
\us\ Catalog was our best option, even with its known defects. Because
we have used the \us\ Catalog as the input for the tesellation
process, our DBF catalogue was affected by the defects of the
\us\ Catalog. Although we have made an effort to minimise the damage,
we cannot completely guarantee the absence of misidentification in our
final DBF catalogue. So, the user should confirm the validity of the
DBFs before using them.

\begin{figure}
  \centering
  \includegraphics[width=\columnwidth]{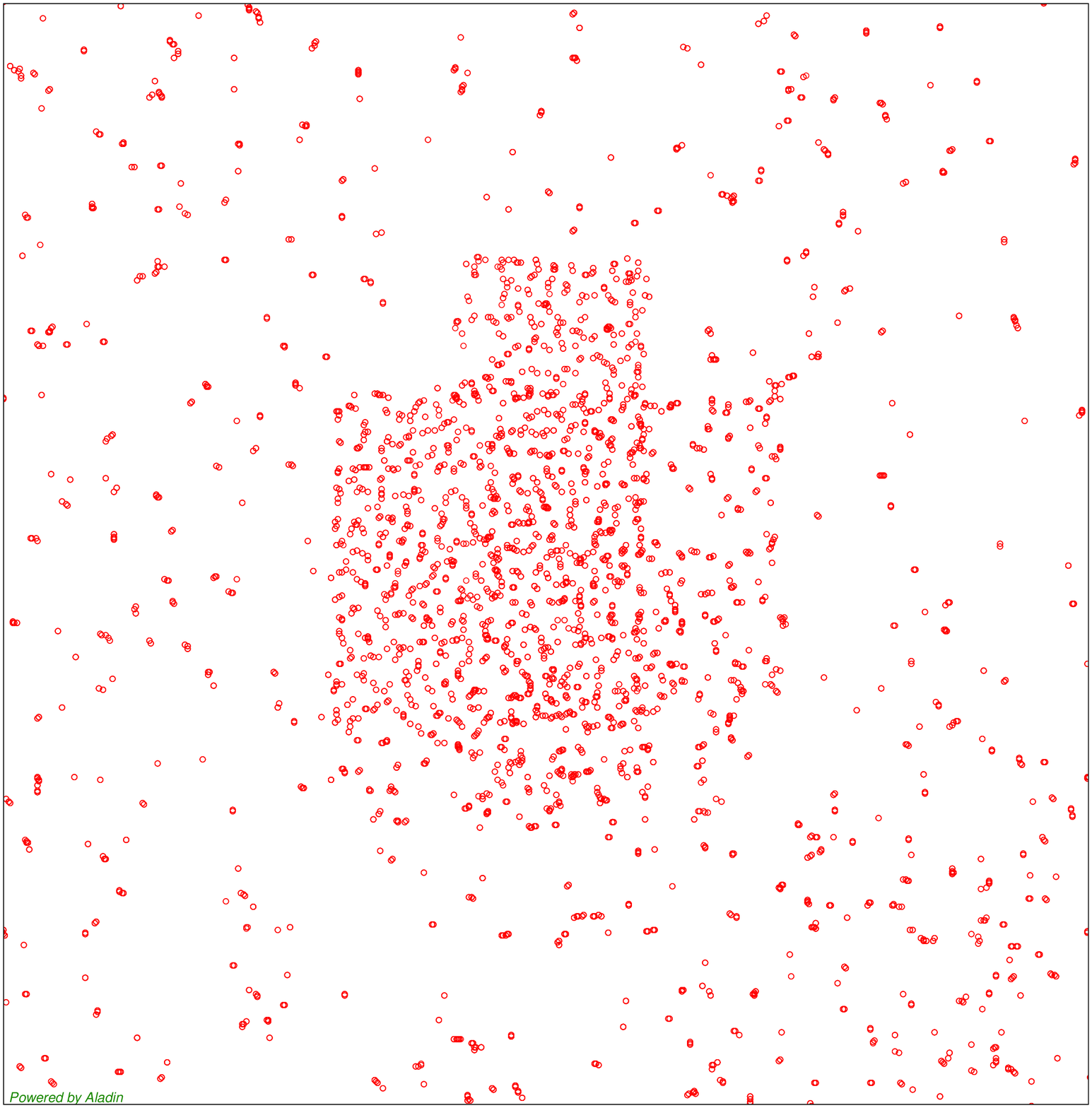}\hfill\\
  \caption{Abnormal concentration of DBFs with $m_\dd{th}$\,=\,15
    in the region of the sky of 12\arcdeg$\times$12\arcdeg centered at
    ($\alpha_{J2000}$,$\delta_{J2000}$)\,=\,(30\arcdeg,00\arcdeg).}
  \label{defect1}
\end{figure}
 
\begin{figure}
  \centering
  \includegraphics[width=\columnwidth]{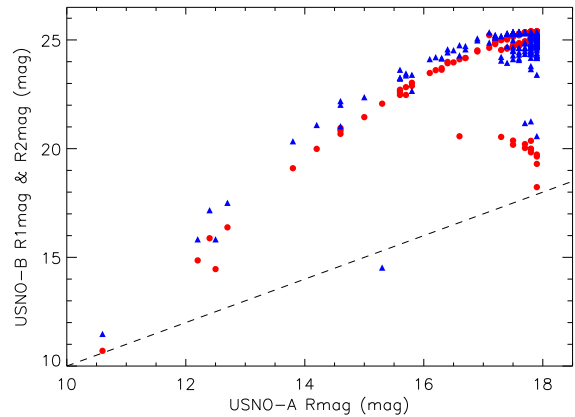}\hfill\\
  \caption{Comparison between USNO-A {\it Rmag} and \us\ {\it R1mag}
    (red circles) and {\it R2mag} (blue triangles) magnitudes of the
    common objects in region of the sky centered at
    ($\alpha_{J2000}$,$\delta_{J2000}$)\,=\,(31.9\arcdeg,--37.1\arcdeg)
    and with a radius if 8.5\arcmin. Dashed line depicts the
    one-to-one correlation.}
  \label{USNO_AvsB}
\end{figure}

During the first inspection of the sky distribution of the DBFs found,
we realized that there were abnormal concentrations of DBFs in two
rectangular-shaped regions, one around
($\alpha_{J2000}$,$\delta_{J2000}$)\,=\,(30\arcdeg,00\arcdeg) and the
other around (27\arcdeg,--37\arcdeg); see Fig.\,\ref{defect1}. This
high density of DBFs reflects an abnormally high value of the
$m_\dd{R}$ in the \us\ Catalog, probably because of a wrong
photometric calibration of the photographic plates. In
Fig.\,\ref{USNO_AvsB} we show a comparison between USNO-A {\it Rmag}
and \us\ {\it R1mag} and {\it R2mag} of the objects in common in a
small region of the sky where abnormally high density of DBFs was
found. The values of \us\ {\it R1mag} and {\it R2mag} are usually more
than 7 mag higher than those of USNO-A {\it Rmag}. Thus, the DBFs
found on this region were clearly unreliable. Then, we defined four
rectangular regions (see Table\,\ref{USNOA}) covering the areas with
an abnormally high density of DBFs and used the USNO-A Catalog in the
tessellation instead of \us\ Catalog.

\begin{table}
   \caption[]{Regions where USNO-A Catalog was used instead of \us
     Catalog.}
     \label{USNOA}
   \begin{tabular}{cccc}
     \hline
     \hline
     \noalign{\smallskip}
      \multicolumn{2}{c}{RA (\arcdeg)}  &  \multicolumn{2}{c}{DEC (\arcdeg)} \\
      \multicolumn{4}{c}{(J2000)} \\
      \ccol{min} & \ccol{max}     & \ccol{min} & \ccol{max} \\
      \noalign{\smallskip} \hline \noalign{\smallskip}
       21.75  & 32.70 & --37.86 & --32.91 \\
       23.25  & 28.95 & --42.46 & --37.75 \\
       28.95  & 31.20 &  +2.01  &  +3.56 \\
       27.64  & 32.45 & --2.65  &  +2.09 \\
       \noalign{\smallskip}
       \hline
   \end{tabular}
\end{table}

\cite{Barron08} have found that $\sim$2.3 per cent of the \us\ Catalog
entries do not correspond to real astronomical sources; they are
mainly - but not only - spurious entries, caused by diffraction spikes
and circular reflection haloes around bright stars in the original
imaging data (see figs\,1, and 2 of their paper). In our search for
DBFS, the effect of these spurious entries is clear; some of the DBFs
might not have been found during the process because of the presence
of one or more spurious sources in the field. Thus, these spurious
entries might have decreased the number of DBFs, but they have never
introduced unreal DBF detections.

\begin{figure}
  \centering
  \includegraphics[width=\columnwidth]{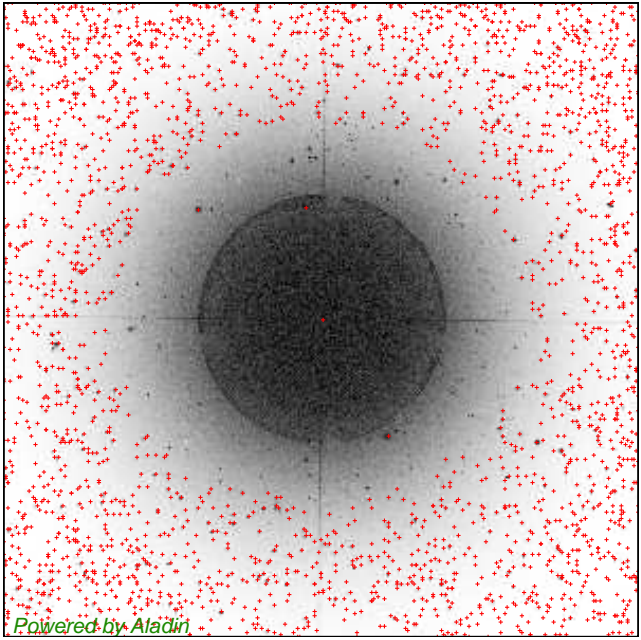}\hfill\\
  \caption{\us\ Catalog entries superimposed to the DSS-red
    30\arcmin$\times$30\arcmin\ image centered at Vega.}
  \label{defect2}
\end{figure}

\begin{figure}
  \centering
  \includegraphics[width=\columnwidth]{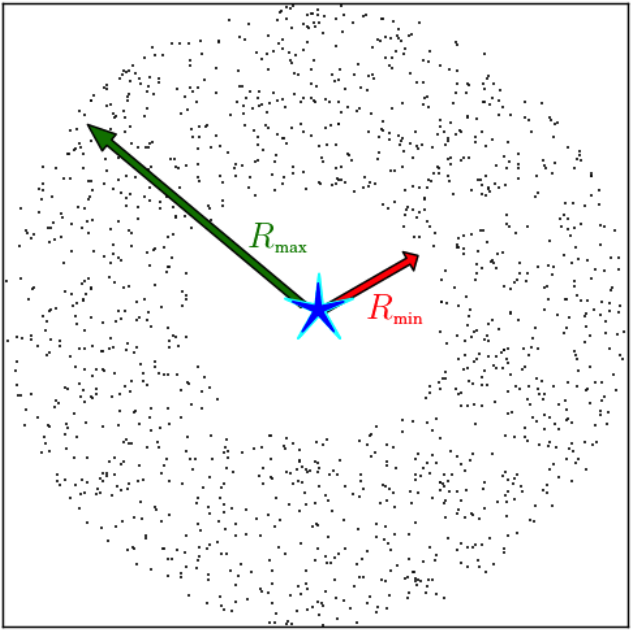}\hfill\\
  \caption{Illustration of the parameters defined for the automatic
    determination of the radii around bright stars, which are devoid
    of neighbouring (fainter) stars. The parameter sought is
    $R_{\mbox{\scriptsize min}}$, where $R_{\mbox{\scriptsize max}}$
    is the maximum region around the bright object where star density
    is determined. The basic hypothesis of the method is the
    assumption that the stellar density in the annulus
    between $R_{\mbox{\scriptsize min}}$ and $R_{\mbox{\scriptsize
        max}}$ is roughly constant.}
  \label{mide_circles}
\end{figure}

\begin{figure}
  \centering
  \includegraphics[width=\columnwidth]{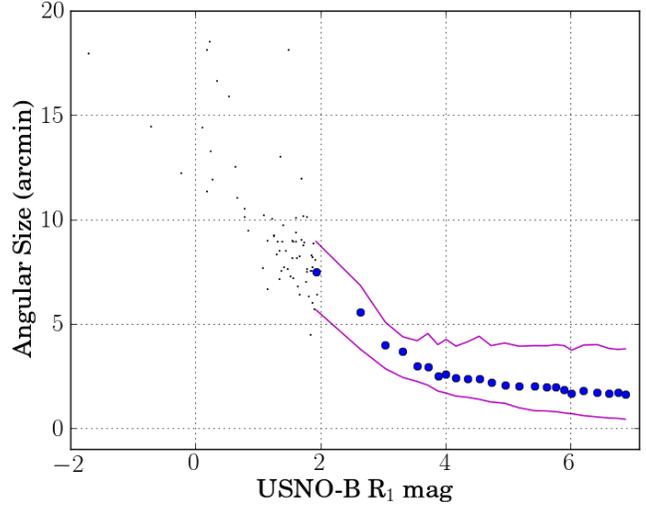}\hfill\\
  \caption{Variation of the size of the artificial void-of-stars
    regions with the \us\ {\it R1mag}. For stars brighter than
    \mbox{$m_\dd{R}=2$ mag} (small black dots), the radius devoid of
    neighbouring stars is shown for each individual star. For stars
    fainter that \mbox{$m_\dd{R}=2$ mag}, the median of the radii are
    plotted as blue filled circles. The magenta lines indicate the
    percentiles 15.9 and 84.1 (which approximately correspond to $\pm
    1\sigma$ confidence limits in a normal distribution).}
  \label{sizes}
\end{figure}

However, there are other defects in the \us\ Catalog that could affect
our tessellation process in a worse way. In Fig.\,\ref{defect2}, we
show the \us\ Catalog entries around Vega, which is one of the
brightest stars of our sky. The brightest stars ($<$7 mag) are
surrounded by a circular halo in the plate images, caused by internal
reflections in the camera. Consequently, the regions around the
brightest stars in the \us\ Catalog are artificial regions that are
void-of-stars, and in many cases these have been wrongly identified as
DBFs.

With the help of the scripting capabilities of Aladin, we have
estimated the radius of the region devoid of neighbour stars around
each of the \mbox{$25\,406$} \us\ stars down to \mbox{$m_\dd{R}=7.0$
  mag}. For that purpose (see Fig.\,\ref{mide_circles}), we consider
that around the bright stars there is a region devoid of stars up to a
radius $R_{\mbox{\scriptsize min}}$. The automatic determination of
this parameter is not difficult if we assume that the stellar
density beyond this radius, and up to $R_{\mbox{\scriptsize max}}$
(which, in our case was set to 30\arcmin), is constant. Under this
circumstance, it is easy to show that the number of stars within a
radius $r$ (with $r > R_{\mbox{\scriptsize min}}$) is given by
\begin{equation}
N(<r) = N_{\mbox{\scriptsize tot}} \; \frac{r^2-R_{\mbox{\scriptsize min}}^2}
{R_{\mbox{\scriptsize max}}^2-R_{\mbox{\scriptsize min}}^2},
\label{eq_nr}
\end{equation}
where $N_{\mbox{\scriptsize tot}}$ is the total number of neighbouring
stars in the annulus between $R_{\mbox{\scriptsize min}}$ and
$R_{\mbox{\scriptsize max}}$. Because we were interested in obtaining
an automatic and robust estimation of $R_{\mbox{\scriptsize min}}$, we
have followed an approach based on the incomplete method of Thail (see
e.g.\ \citealt{DeMuth06}). As with most non-parametric procedures, all
the objects in the field (within $R_{\mbox{\scriptsize max}}$) were
first sorted in ascending order using the distance from the central
bright star. This provided a collection of $N_{\mbox{\scriptsize
    tot}}$ points of coordinates $(R_i,N_i)$ where
$i=1,2,3,....,N_{\mbox{\scriptsize tot}}$. Here, $R_i$ is the distance
from the centre and $N_i$ is the number of objects within the circle of
radius $R_i$. From any pair of such points, and considering
equation~(\ref{eq_nr}), it is possible to obtain an estimation of
$R_{\mbox{\scriptsize min}}$ as
\begin{equation}
R_{\mbox{\scriptsize min}}^2 = \frac{N_b \; R_a^2-N_a \;  R_b^2}{N_b-N_a},
\end{equation}
where $R_{\mbox{\scriptsize min}} \leq R_a < R_b \leq
R_{\mbox{\scriptsize max}}$. In this way, it is possible to compute
$N_{\mbox{\scriptsize tot}}/2$ values for $R_{\mbox{\scriptsize min}}$
by using the last equation on pairs of points where
$a=1,2,3,...,N_{\mbox{\scriptsize tot}}/2$ and
$b=a+N_{\mbox{\scriptsize tot}}/2$ (in fact, this is the basics of the
Thail method). Finally, we have determined the median of the
$N_{\mbox{\scriptsize tot}}/2$ estimates of $R_{\mbox{\scriptsize
    min}}$.

In Fig.\,\ref{sizes}, we show the variation of the size of the region
devoid of neighbouring stars, computed inthe way just explained, with
$m_\dd{R}$. On the basis of this diagram, we applied the following
conservative criterion to avoid false DBFs in our catalogue: we have
removed all DBFs whose centres were closer than 25\arcmin to stars
with $m_\dd{R}$\,$\le$\,0.5, 15\arcmin to stars with
0.5\,$<$\,$m_\dd{R}$\,$\le$\,3.0, and 10\arcmin to stars with
0.3\,$<$\,$m_\dd{R}$\,$\le$\,6.0. This criterion was not applied to
the DBF with 18 $m_\dd{th}$, because they were all visually
inspected. The percentages of removed DBFs are 1, 2, and 9\% for 15,
16, and 17 $m_\dd{th}$, respectively. As expected, the percentage
increases with $m_\dd{th}$ because the number of wrong DBFs should
be similar for all catalogues.

\begin{figure}
  \centering
  \includegraphics[width=\columnwidth]{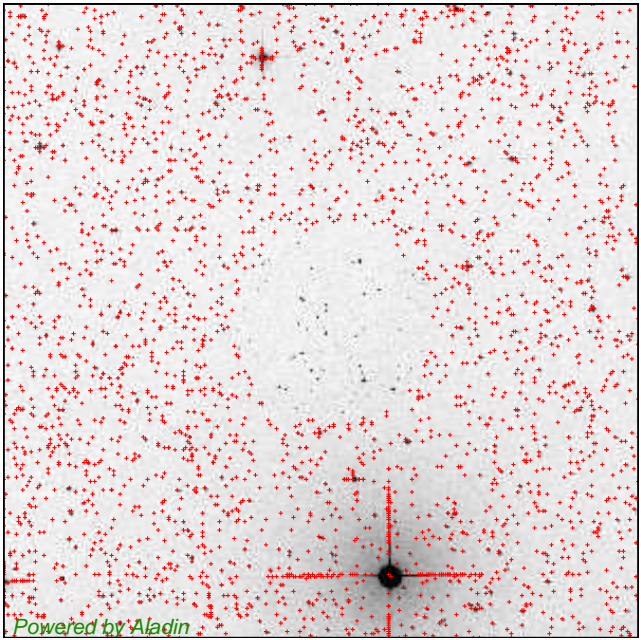}\hfill\\
  \caption{\us\ Catalog entries superimposed to the DSS-blue
    30\arcmin$\times$30\arcmin image centered at
    ($\alpha_{J2000}$,$\delta_{J2000}$)\,=\,(111.689\arcdeg,$-$78.893\arcdeg).}
  \label{defect3}
\end{figure}

In Fig.\,\ref{defect3}, we show another defect of the
\us\ Catalog. There are no entries in a circular region of
$\sim$9\arcmin diameter around the position in the sky
($\alpha_{J2000}$,$\delta_{J2000}$)\,=\,(07:26:40,$-$78:53:50). Our
tessellation process detected this region as void of star and it has
included a corresponding BF entry in the catalogue. The same problem
occurred with another region of $\sim$16\arcmin diameter around
($\alpha_{J2000}$,$\delta_{J2000}$)\,=\,(20:12:40,$-$74:14:39). These
two false void-of-star regions were detected by a change in the visual
inspection of the DBFs, so we cannot discard further similar defects
in the \us\ Catalog, and consequently the possible presence of
spurious BFs in our catalogue. These two detected spurious entries
were manually rejected from our catalogue.

\subsection{Results}
  \label{section:result}

The final result is a catalogue that is currently the deepest
catalogue of regions that are void of stars. The catalogue is
accessible through the TESELA tool (as explained in
Section\,\ref{section:TESELA}).

Table~\ref{results} presents the quantitative description of the above
results. Column 1 gaves the threshold magnitude $m_\dd{th}$ and column
2 lists the number of DBF regions found $N_\dd{DBF}$. Columns 3, 4,
and 5 indicate the mean radius $\overline\rho$, the standard deviation
$\sigma_{\rho}$, and the maximum DBF radius $\rho_\dd{max}$,
respectively. The integrated mean flux $\overline{F_{\dd{int}}}$, the
standard deviation $\sigma_{F_\dd{int}}$, and the minimum integrated
flux $F_\dd{min}$ and maximum integrated flux $F_\dd{min}$ are given
in columns 6, 7, 8, and 9, respectively.

\begin{table*}
   \caption[]{Results of the tessellation process using the
     \us\ Catalog with the {\it R} filter.}
     \label{results}
   \begin{tabular}{rrrrrrrrr}
     \hline
     \hline
     \ccol{(1)}     & \ccol{(2)}         & \ccol{(3)}               & \ccol{(4)}             & \ccol{(5)}            & \ccol{(6)}          & \ccol{(7)}                 & \ccol{(8)}            & \ccol{(9)}          \\
     \noalign{\smallskip}                                                                                                                                                                                              
     \ccol{$m_\dd{th}$} & \ccol{$N_\dd{DBF}$} & \ccol{$\overline{\rho}$} & \ccol{$\sigma_{\rho}$} & \ccol{$\rho_\dd{max}$} & \ccol{$\overline{F_{\dd{int}}}$} & \ccol{$\sigma_{F_\dd{int}}$}  & \ccol{$F_\dd{min}$} & \ccol{$F_\dd{max}$}  \\
     \ccol{(mag)}   &                    & (\arcmin)                & (\arcmin)             &   (\arcmin)            &  (erg/s/cm$^{2}$/A) &  (erg/s/cm$^{2}$/A)         &   (erg/s/cm$^{2}$/A)&   (erg/s/cm$^{2}$/A)    \\

      \noalign{\smallskip} \hline \noalign{\smallskip}
    $^{a}$11.0   & 1.477.348  & 10    &   4 & 51.2 & \\
          15.0   &    77.277 &  5.5 & 0.5 & 14.2 & 1.6e-14 & 7.0e-15 & 2.3e-17 & 2.2e-13 \\
          16.0   &     6.062 &  5.4 & 0.5 & 12.8 & 8.5e-15 & 3.0e-15 & 3.1e-16 & 3.2e-14 \\
          17.0   &       332 &  5.6 & 0.7 &  9.0 & 3.8e-15 & 1.9e-15 & 1.5e-16 & 1.2e-14 \\
          18.0   &        74 &  5.8 & 0.8 &  9.0 & 1.8e-15 & 9.3e-16 & 3.1e-16 & 4.8e-15 \\
   \noalign{\smallskip}
   \hline
   \end{tabular}
   \begin{list}{}{}
      \item[$^{a}$] From the 11 threshold magnitude catalogue of
        Paper~I obtained with the combination of Tycho-2 {\it B} and
        {\it V} bands.
   \end{list}
\end{table*}

\begin{figure*}
\centering
\includegraphics[width=\columnwidth]{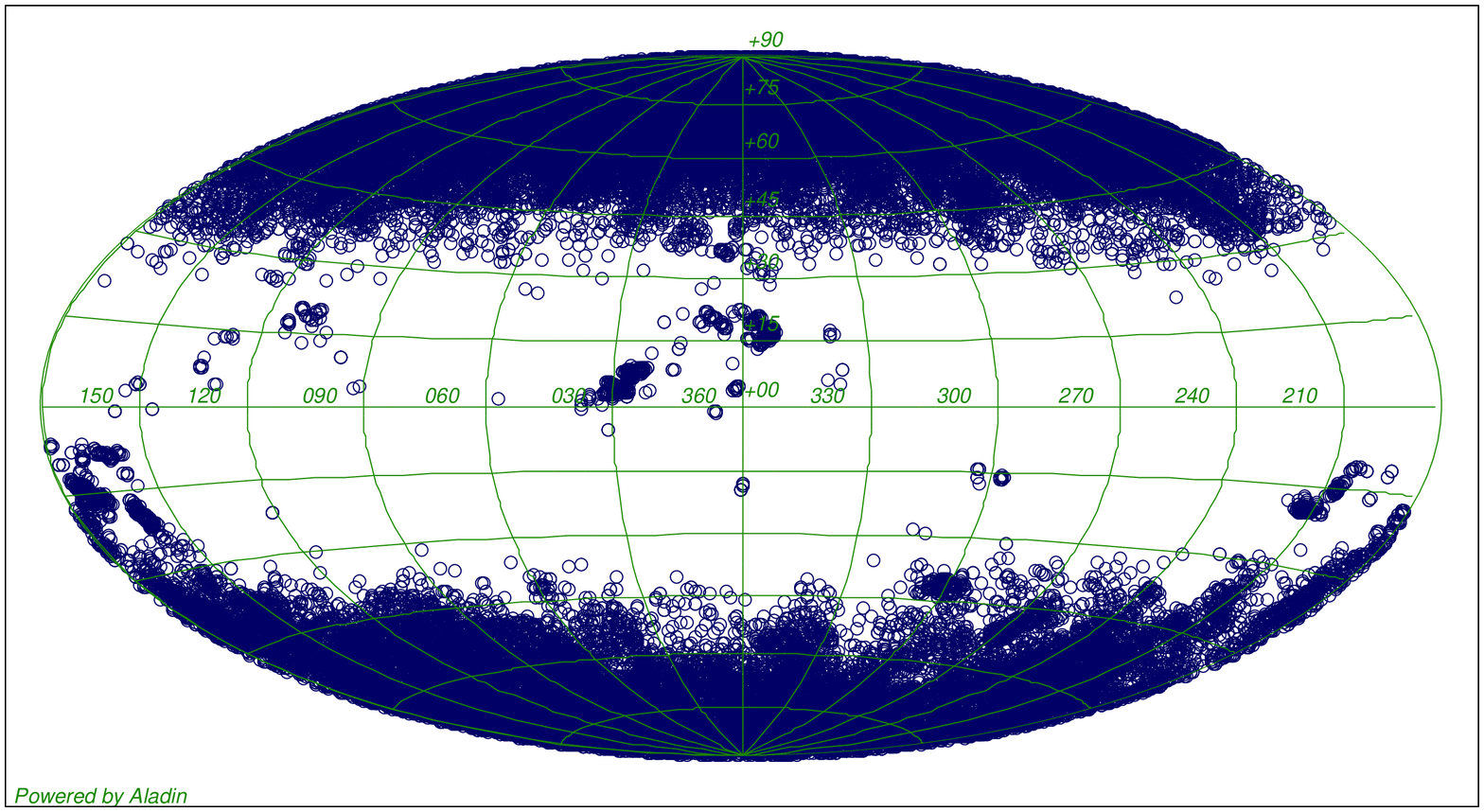}
\includegraphics[width=\columnwidth]{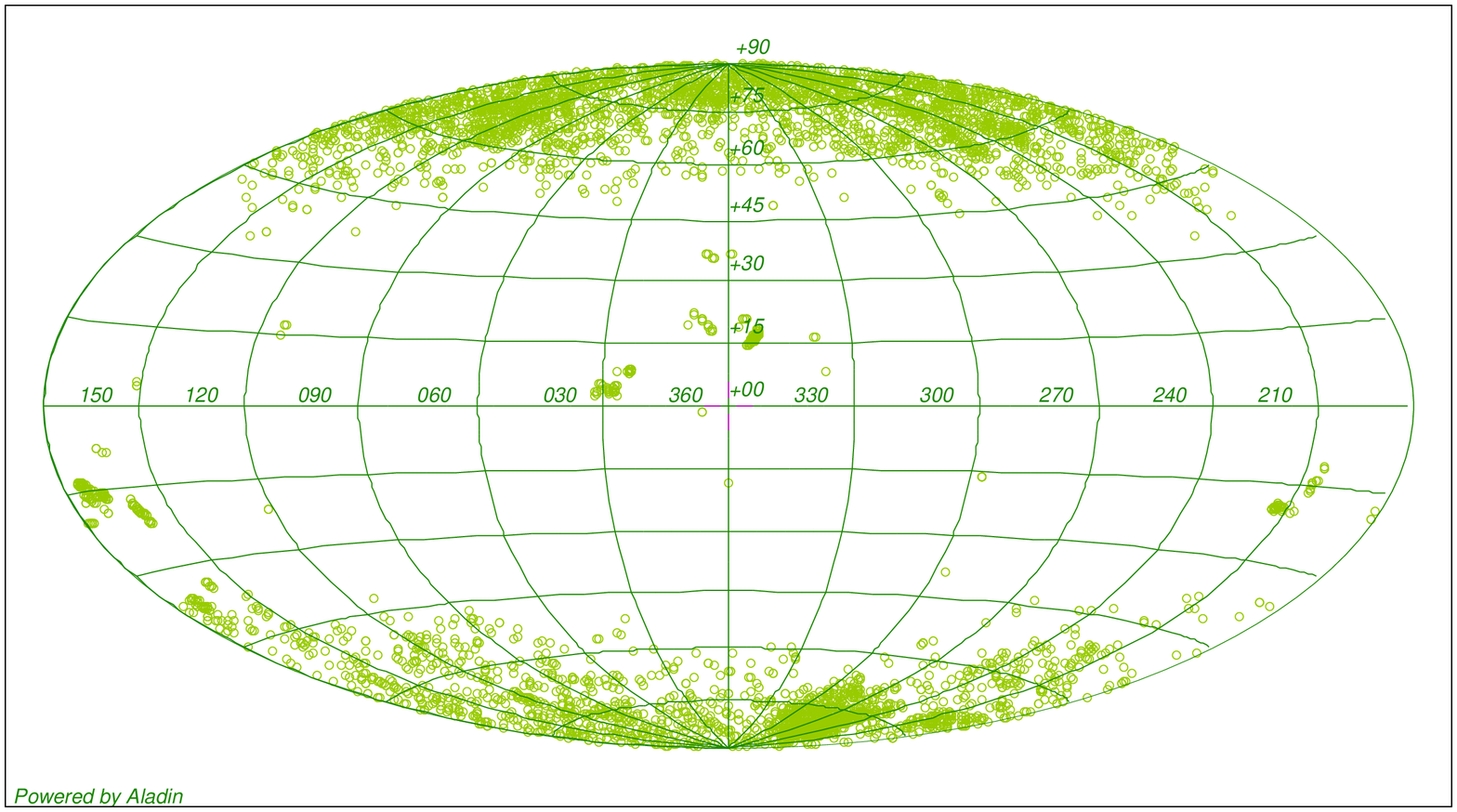}
\vskip 1mm
\includegraphics[width=\columnwidth]{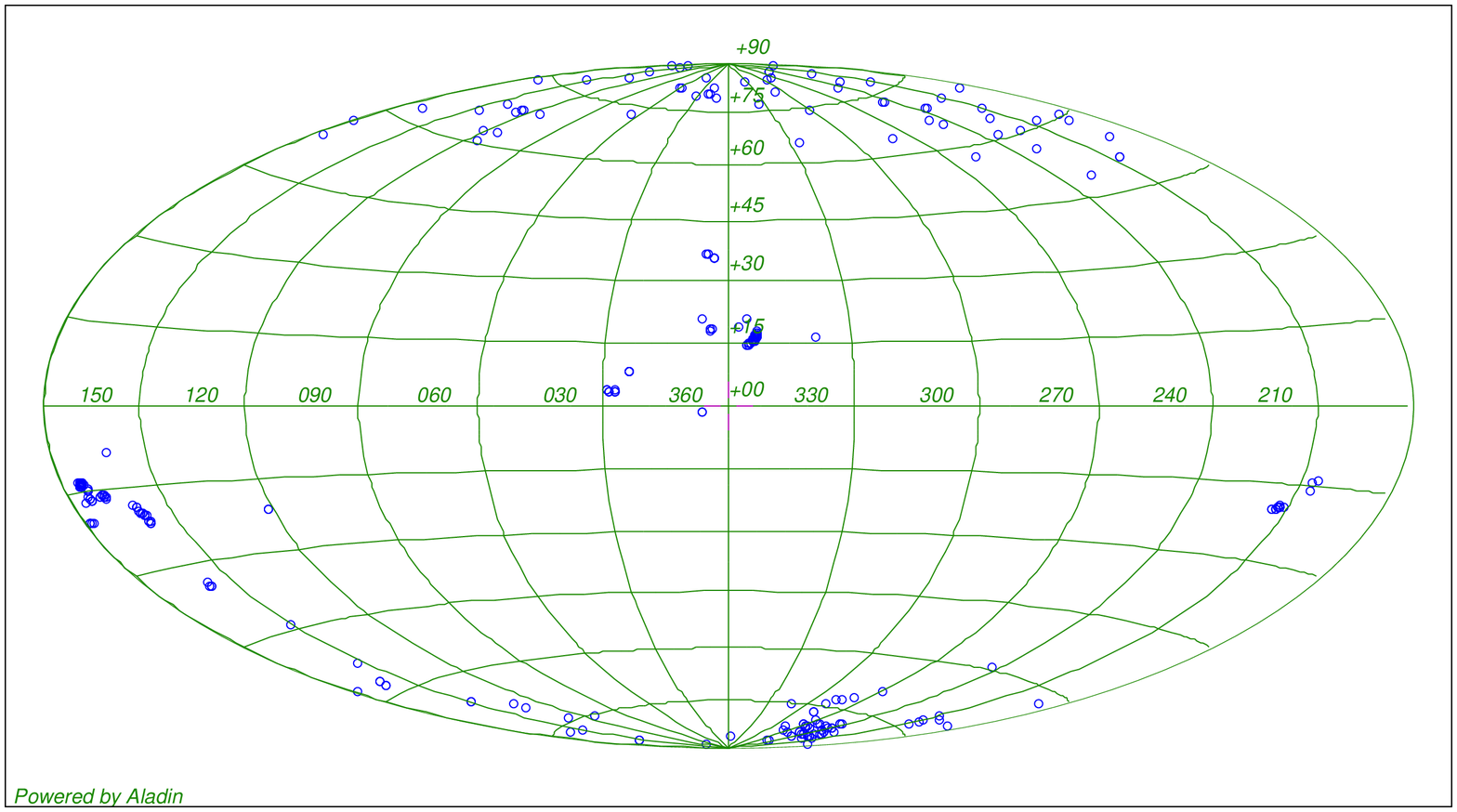}
\includegraphics[width=\columnwidth]{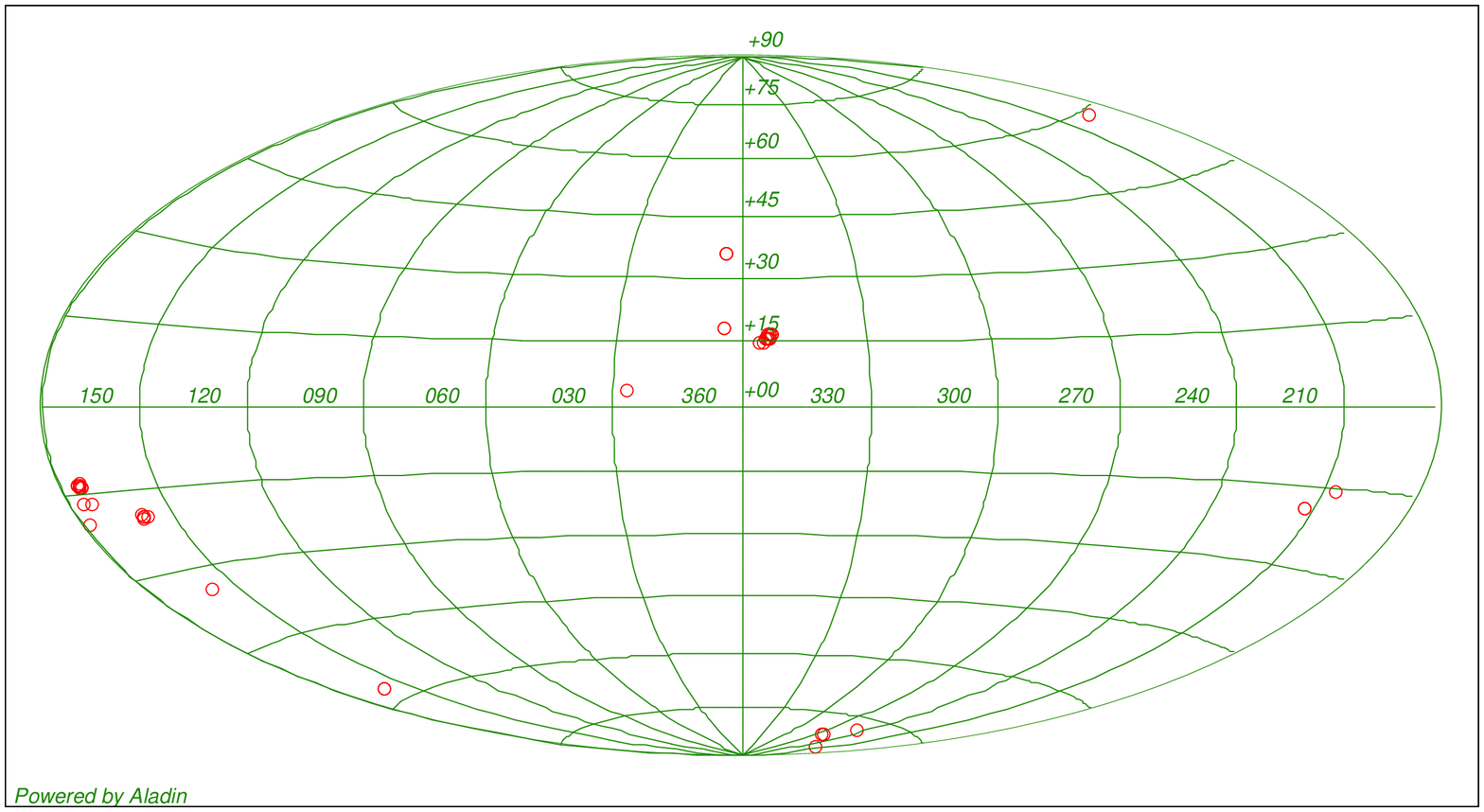}
\caption{Galactic positions of the DBFs with $m_\dd{th}$ 15, 16, 17,
  and 18 in the upper-left, upper-right, lower-left, and lower-right
  panel, respectively.}
\label{maps}
\end{figure*}

In Fig.\,\ref{maps}, we show the maps of the DBFs found in galactic
coordinates (IAU 1958). As expected, the major concentration of DBFs
is close to the galactic poles, with a low number of DBFs close to
the galactic plane.

In order to provide a short list of DBFs that can be used at the
telescope, we have selected, among the best regions found, one DBF for
approximately a half hour in RA in each hemisphere, as provided in the
Table\,\ref{northDBFs}. The equatorial and galactic coordinates are
given first (columns 1--4). The radius $\rho$ of the DBF is given in
column 5. The number of stars $N_{*}$ with $m_\dd{R}$ lower than 17.5
and 22.0 mag are listed in columns 6 and 7, respectively. Columns 8
and 9 show $m_\dd{R}$ of the faintest and brightest star in the
field. The integrated flux $F_{\dd{int}}$ is in column 10. The last
four columns 11--14 give the galactic extincition $E(B-V)$ as
described below. There are some regions at the Southern hemisphere
where no DBFs have been found. Some of these DBFs have been validated
with the GTC and are they are regularly used in the nightly operation
of the telescope (see Section\,\ref{section:GTC}).

Although our main goal in this work was to provide a list of DBFs for
data calibration, the DBFs found can also be used for other
interesting purposes (e.g. deep pencil-beam extragalactic surveys). In
this sense, the galactic extinction in the direction of the sky where
the DBFs are located is useful information that would help in the
identification of suitable DBFs. Thus, we used The Galactic Dust
Reddening and Extinction utility of the NASA/ IPAC Infrared Science
Archive\footnote{http://irsa.ipac.caltech.edu/applications/DUST} to
estimate the galactic extinction from \cite{Schlegel98} for each DBF
in our catalogue. In this way, five new parameters were added to the
catalogue: $E(B-V)$ at the center of the DBF, the average $E(B-V)$ in
the sky region defined by the DBF, together with its standard
deviation, and the maximum and minimum $E(B-V)$ in this sky area.


\section{Deep blank fields for OSIRIS/GTC}
 \label{section:GTC}

We have used the 10.4m GTC, located at Observatorio Roque de los
Muchachos in La Palma (ORM), Canary Islands, Spain, as a test-bed for
the validation of our DBFs. The telescope is currently equipped for
scientific exploitation with the OSIRIS optical facility, which
includes imaging with broad-band filters from the Sloan set ($ugriz$),
as well as a very flexible set of narrow-band filters that can be used
when requested.

The OSIRIS instrument consists of a mosaic of two Marconi CCD
detectors, each with 2048\,$\times$\,4096 pixels and a total
unviggnetted field of view of 7.8\,$\times$\,7.8\arcmin, giving a
plate scale of 0.127\arcsec/pix. For this reason, the series of BFs
requested for the nightly operation of the instrument must be at least
10\arcmin wide, it must be larger than the OSIRIS field of view, and
it must also take into account the dithering between consecutive
exposures that is usually performed to eliminate the contribution of
the stars in producing the final flat-field image.

To date, the most comprehensive list of BFs available for observations
carried out at the ORM has been the collection made by
Azzaro. However, the large aperture of the GTC means that these fields
(which are empty up to magnitudes 10--16) are too shallow to be used
as BFs for the GTC, because they appear relatively full of stars once
an OSIRIS/GTC image is obtained.

\begin{figure*}
  \includegraphics[width=17.5cm]{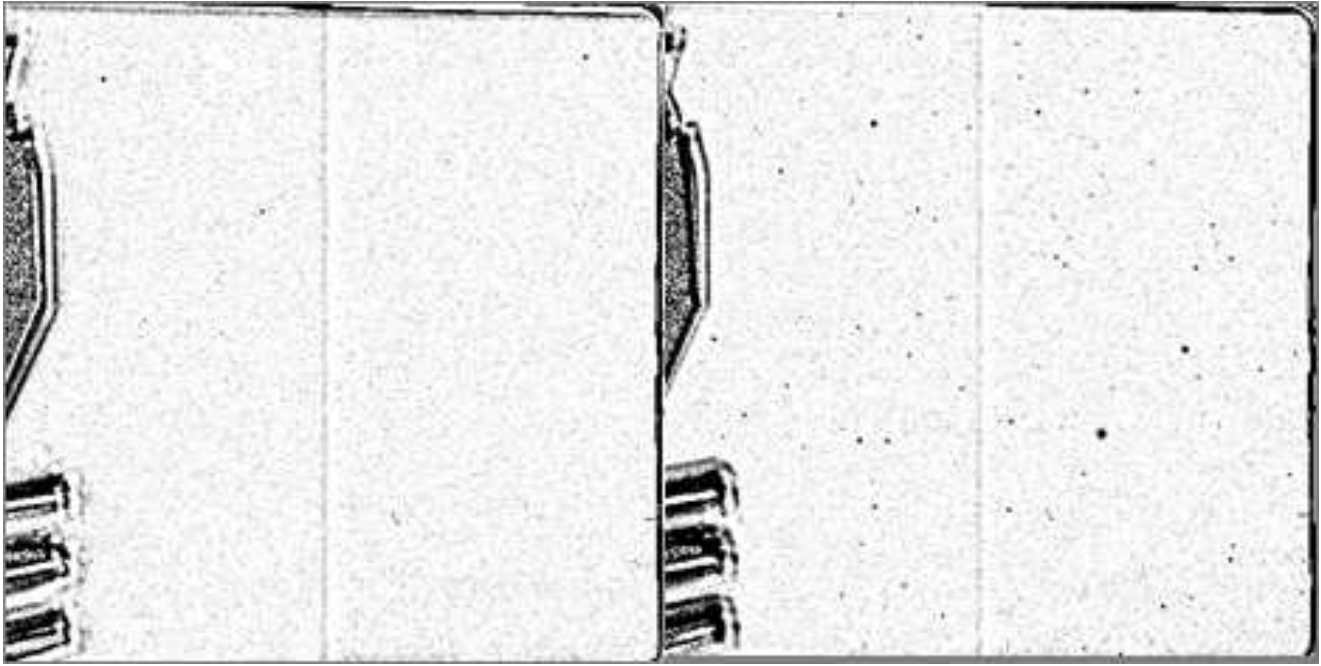}

  \caption{Comparison of two images taken with OSIRIS. They are 2-s
    exposure images at the Sloan $r$ filter, with a background level
    of $\sim$30,000 ADUs, of a DBF from our catalogue (\emph{left
      panel}) and a BF from Azzaro catalogue (\emph{right panel}) in
    the same region of the sky. The size of both fields is
    8.5\arcmin$\times$8.5\arcmin. There is an evident improvement of
    our DBF with the one from the Azzaro catalogue. Note that the two
    apparent detections, both in the right-upper part in CCD1 (left
    side of the detector) and in left-bottom part in CCD2 (right side
    of the detector), are dust grain effects on the OSIRIS CCDs, and
    they do not correspond with any star in the field (they are also
    visible in the Azzaro field).}
  \label{BFs}
\end{figure*}

During the nightly operation of OSIRIS/GTC, an automatic script is
used to obtain a series of broad-band flat-field images in Sloan
$griz$ filters (Sloan $u$ filter is of rare use). Because of the
particularities of the OSIRIS shutter, in order to obtain a
photometric accuracy better than 0.1\% in the images, exposure times
of the order of 1--2 s are needed. To define the limiting magnitude
for the BFs to be used at the GTC, a complete set of Azzaro fields was
observed over nearly a year of operation at the telescope, and were
reduced photometrically to determine the magnitude of the detected
stars under good photometric conditions. For 1--2 s of exposure time,
in Sloan $r$ filter, and with nearly 30\,000 ADUs of sky background
level (which represents half of the full well of the detector), stars
as faint as 17.5--18.0 mag (depending on the seeing conditions) were
detected. This is the fainter limit that we have imposed on the BFs to
be used with OSIRIS/GTC, which might be empty regions of the sky up to
$r'=17.5$ mag. This is impossible to achieve with the Azzaro
catalogue. In addition, for operating reasons, those BFs should
provide a complete RA coverage. These two constraints are very
restrictive, but they can sometimes be met using our DBF
catalogue. Thus, we have been able to create a list of 87 DBFs that
can be used with OSIRIS/GTC, although it was not always possible to
archive the $m_\dd{th}$ limit. This list provides at least one of
these DBFs both at the beginning or at the end of every night of the
year, which is extremely useful for the calibration of the scientific
data delivered by OSIRIS instrument. This list has superseded the
Azzaro catalogue in the nightly operation of the telescope. Some of
the DBFs selected can be found in Tables\,\ref{northDBFs}. There is an
evident advantage to using the new DBFs rather than those from the
Azzaro collection, because they have been shown to be much better. In
Fig.\,\ref{BFs}, we show a comparison of one of the DBFs used at the
GTC with another one from the Azzaro catalogue.

\section{TESELA}
 \label{section:TESELA}

In order to provide easy access to the DBF catalogue, we have
incorporated it into TESELA, a WEB accessible tool, developed within
the Spanish Virtual Observatory\footnote{\tt
  http://svo.laeff.inta.es}, which can be publicly accessed from {\tt
  http://sdc.cab.inta-csic.es/tesela.}

The tool was implemented with a data base containing the new computed
DBF regions from the \us\ Catalog. Thus, through an user-friendly
interface, the user can now perform a cone-search around a position in
the sky in either the BF catalogue from \mbox{Tycho-2} (see Paper~I
for a extended description) or the DBF catalogue from
\us\ Catalog. Through the search form for the DBF catalogue, users can
select the $m_\dd{th}$ and the search radius. The cone-search can be
made in either equatorial or galactic coordinates.

TESELA presents the result of the cone-search in a table containing
the same main properties of the DBFs presented in
Tables\,\ref{northDBFs}. This table can be downloaded in CSV format
for further use. TESELA also provides users with the possibility of
visualizing the data. To do this, TESELA takes advantage of
Aladin. Thanks to this connection with Aladin, we have provided the
user with the full capacity and power of the VO. Using this Aladin
window, users are allowed to load images and catalogues both locally
and from the VO, having full access to the whole universe resident in
the VO. This can be very helpful to determine the potential influence
of relatively bright nebulae and extragalactic sources in the regions
sought.

TESELA sends to Aladin the searching area, which is plotted in the
first plane with a red circle, and the DBFs, which are shown by blue
circles in a second layer. The size of the blue circles corresponds
with the size of the DBFs. Finally, Aladin charges from the VO in a
last layer the objects of NGC~2000.0 (Complete New General Catalogue
and Index Catalogue of Nebulae and Star Clusters;
\citealt{Sinnott97}). Note that, for obvious reasons, Solar system
objects have been not considered, and they must be taken into account
in order to make use of BF regions close to the ecliptic at a given
date.

\section*{Acknowledgements}

This work was partially funded by the Spanish MICINN under the
Consolider-Ingenio 2010 Program grant CSD2006-00070: First Science
with the GTC\footnote{\tt http://www.iac.es/consolider-ingenio-gtc}.
This work was also supported by the Spanish Programa Nacional de
Astronom\'{\i}a y Astrof\'{\i}sica under grants AyA2011-24052 and
AYA2009-10368, and by AstroMadrid\footnote{\tt
  http://www.astromadrid.es} under project CAM S2009/ESP-1496. This
work is based on observations made with the GTC, installed in the
Spanish Observatorio del Roque de los Muchachos of the Instituto de
Astrofísica de Canarias, in the island of La Palma. This work has made
use of {\sc ALADIN} developed at the Centre de Donne\'{e}s
Astronomiques de Strasbourg, France. This research has made use of the
NASA/ IPAC Infrared Science Archive, which is operated by the Jet
Propulsion Laboratory, California Institute of Technology, under
contract with the National Aeronautics and Space Administration.


\scriptsize

\begin{landscape}
\onecolumn

\begin{longtable}{rcrccrrccccccc}
  \caption[]{Short list of selected DBFs.}
  \label{northDBFs}\\

  \hline \hline \noalign{\smallskip}
  \multicolumn{1}{c}{(1)} & 
  \multicolumn{1}{c}{(2)} & 
  \multicolumn{1}{c}{(3)} & 
  \multicolumn{1}{c}{(4)} & 
  \multicolumn{1}{c}{(5)} & 
  \multicolumn{1}{c}{(6)} & 
  \multicolumn{1}{c}{(7)} & 
  \multicolumn{1}{c}{(8)} & 
  \multicolumn{1}{c}{(9)} & 
  \multicolumn{1}{c}{(10)} & 
  \multicolumn{1}{c}{(11)} & 
  \multicolumn{1}{c}{(12)} & 
  \multicolumn{1}{c}{(13)} & 
  \multicolumn{1}{c}{(14)} \\
  \multicolumn{1}{c}{RA} &
  \multicolumn{1}{c}{DEC} & 
  \multicolumn{1}{c}{Gal Long} & 
  \multicolumn{1}{c}{Gal Lat} & 
  \multicolumn{1}{c}{$\rho$} & 
  \multicolumn{2}{c}{$N_{*}$ with $m_\dd{R}$} & 
  \multicolumn{2}{c}{$m_\dd{R}$} & 
  \multicolumn{1}{c}{$F_{\dd{int}}$} & 
  \multicolumn{4}{c}{E(B-V) (mag)} \\
  \multicolumn{1}{c}{(J2000)} & 
  \multicolumn{1}{c}{(J2000)} & 
  \multicolumn{1}{c}{(\arcdeg)} & 
  \multicolumn{1}{c}{(\arcdeg)} & 
  \multicolumn{1}{c}{(\arcmin)} &  
  \multicolumn{1}{c}{$<$\,17.5} & 
  \multicolumn{1}{c}{$<$\,22.0} &
  \multicolumn{1}{c}{Max} &
  \multicolumn{1}{c}{Min} & 
  \multicolumn{1}{c}{(erg/s/cm$^2$/A)} & 
  \multicolumn{1}{c}{Center} & 
  \multicolumn{1}{c}{Mean} & 
  \multicolumn{1}{c}{Max.} & 
  \multicolumn{1}{c}{Min.} \\
  \noalign{\smallskip} \hline \noalign{\smallskip}
\endfirsthead

\caption{continued.}\\
  \hline \hline \noalign{\smallskip}
  \multicolumn{1}{c}{(1)} & 
  \multicolumn{1}{c}{(2)} & 
  \multicolumn{1}{c}{(3)} & 
  \multicolumn{1}{c}{(4)} & 
  \multicolumn{1}{c}{(5)} & 
  \multicolumn{1}{c}{(6)} & 
  \multicolumn{1}{c}{(7)} & 
  \multicolumn{1}{c}{(8)} & 
  \multicolumn{1}{c}{(9)} & 
  \multicolumn{1}{c}{(10)} & 
  \multicolumn{1}{c}{(11)} & 
  \multicolumn{1}{c}{(12)} & 
  \multicolumn{1}{c}{(13)} & 
  \multicolumn{1}{c}{(14)} \\
  \multicolumn{1}{c}{RA} &
  \multicolumn{1}{c}{DEC} & 
  \multicolumn{1}{c}{Gal Long} & 
  \multicolumn{1}{c}{Gal Lat} & 
  \multicolumn{1}{c}{$\rho$} & 
  \multicolumn{2}{c}{$N_{*}$ with $m_\dd{R}$} & 
  \multicolumn{2}{c}{$m_\dd{R}$} & 
  \multicolumn{1}{c}{$F_{\dd{int}}$} & 
  \multicolumn{4}{c}{E(B-V) (mag)} \\
  \multicolumn{1}{c}{(J2000)} & 
  \multicolumn{1}{c}{(J2000)} & 
  \multicolumn{1}{c}{(\arcdeg)} & 
  \multicolumn{1}{c}{(\arcdeg)} & 
  \multicolumn{1}{c}{(\arcmin)} &  
  \multicolumn{1}{c}{$<$\,17.5} & 
  \multicolumn{1}{c}{$<$\,22.0} &
  \multicolumn{1}{c}{Max} &
  \multicolumn{1}{c}{Min} & 
  \multicolumn{1}{c}{(erg/s/cm$^2$/A)} & 
  \multicolumn{1}{c}{Center} & 
  \multicolumn{1}{c}{Mean} & 
  \multicolumn{1}{c}{Max.} & 
  \multicolumn{1}{c}{Min.} \\
  \noalign{\smallskip} \hline \noalign{\smallskip}
\endhead

\hline
\multicolumn{14}{c}{{Continued on Next Page\ldots}} \\
\endfoot

\endlastfoot
 \multicolumn{14}{c}{Northern Hemisphere} \\
00:06:57.17       & +06:06:22.2 & 103.363380 & --55.06648 & 5.03 &   14 & 86  & 20.56 & 16.13 & 8.45E--15 & 0.0738 & (72$\pm$3)E--3    & 0.0771 & 0.0651 \\
00:42:55.80       & +41:20:04.2 & 121.215441 & --21.50889 & 5.19 &   1  &  4  & 19.60 & 17.04 & 4.89E--16 & 0.6767 & (58$\pm$10)E--2   & 0.7419 & 0.4284\\
01:08:41.88       & +06:23:43.5 & 130.659924 & --56.21711 & 5.18 &   5  & 97  & 20.31 & 16.61 & 4.38E--15 & 0.0252 & (260$\pm$14)E--4  & 0.0288 & 0.0230\\
01:35:06.00       & +14:33:23.4 & 138.518359 & --46.98960 & 5.02 &   5  & 155 & 21.01 & 16.45 & 5.63E--15 & 0.0498 & (49$\pm$3)E--3    & 0.0545 & 0.0451\\
02:01:10.30       & +05:06:56.2 & 153.134220 & --53.61670 & 5.08 &   3  & 83  & 19.78 & 17.06 & 3.82E--15 & 0.0425 & (420$\pm$5)E--4   & 0.0428 & 0.0409\\
02:14:26.26       & +14:01:23.9 & 151.573608 & --44.18390 & 5.08 &   4  & 102 & 20.88 & 17.32 & 4.03E--15 & 0.1142 & (113$\pm$7)E--3   & 0.1245 & 0.1007\\
02:56:21.26       & +19:38:41.3 & 159.162056 & --34.29385 & 5.27 &   0  &  17 & 20.16 & 18.46 & 6.24E--16 & 1.9877 & (184$\pm$17)E--2  & 2.0707 & 1.4920\\
$^{a}$03:33:31.70 & +31:02:59.3 & 159.296833 & --20.14533 & 5.03 &   0  &  21 & 19.85 & 18.73 & 5.86E--16 & 2.1341 & (21$\pm$3)E--1     & 2.6959 & 1.5194\\
04:13:56.14       & +28:08:33.0 & 168.217578 & --16.42665 & 5.01 &   1  & 53  & 19.43 & 17.35 & 3.47E--15 & 1.6791 & (169$\pm$6)E--2   & 1.7605 & 1.5595\\
$^{a}$04:26:38.69 & +24:39:00.4 & 172.908158 & --16.71030 & 5.00 &   0  &  12 & 20.03 & 18.88 & 3.12E-16  & 1.1909 & (118$\pm$10)E--2   & 1.3694 & 1.0021\\
04:45:42.65       & +17:11:51.7 & 181.904567 & --17.97142 & 5.28 &   27 & 222 & 20.76 & 16.07 & 2.13E--14 & 0.7981 & (81$\pm$7)E--2    & 0.9400 & 0.6871\\
05:31:50.23       & +12:37:23.5 & 192.273614 & --11.30104 & 5.36 &   24 & 262 & 20.06 & 15.44 & 2.30E--14 & 1.9941 & (20$\pm$3)E--1    & 2.5221 & 1.3798\\
05:47:22.73       & +00:24:17.3 & 205.107319 & --14.02273 & 5.03 &   0  & 37  & 19.55 & 17.53 & 1.83E--15 & 5.9760 & (7$\pm$5)E--0     & 20.533 & 1.8988\\
06:24:09.29       & +84:37:36.8 & 128.888339 &  +26.36465 & 5.19 &   31 & 220 & 20.79 & 15.13 & 2.99E--14 & 0.0694 & (70$\pm$4)E--3   & 0.0772 & 0.0646\\
07:12:48.00       & +56:17:40.6 & 160.555972 &  +25.16972 & 5.14 &   34 & 220 & 20.77 & 15.13 & 3.00E--14 & 0.0538 & (538$\pm$6)E--4  & 0.0550 & 0.0526\\
07:36:47.76       & +69:48:38.5 & 145.747848 &  +29.23289 & 5.34 &   34 & 226 & 20.61 & 15.30 & 2.68E--14 & 0.0289 & (288$\pm$19)E--4 & 0.0320 & 0.0254\\
$^{a}$07:48:36.72 & +41:31:46.9 & 178.046803 &  +27.82759 & 5.36 &   33 & 197 & 20.38 & 15.15 & 2.38E--14 & 0.0439 & (433$\pm$9)E--4   & 0.0456 & 0.0415\\
$^{a}$08:52:36.48 & +24:43:42.2 & 201.164237 &  +36.76278 & 5.22 &   9  & 139 & 20.74 & 16.66 & 8.47E--15 & 0.0323 & (324$\pm$9)E--4   & 0.0338 & 0.0308\\
09:08:21.12       & +74:08:07.4 & 138.975295 &  +35.13226 & 5.05 &   1  & 11  & 20.08 & 16.75 & 1.16E--15 & 0.0374 & (365$\pm$10)E--4 & 0.0379 & 0.0339\\
09:33:13.20       & +28:58:49.1 & 198.410292 &  +46.55147 & 5.05 &   6  & 103 & 19.61 & 16.50 & 5.81E--15 & 0.0178 & (179$\pm$3)E--4  & 0.0184 & 0.0174\\
$^{a}$09:58:19.68 & +24:59:11.4 & 206.046838 &  +51.25283 & 5.20 &   5  & 104 & 19.97 & 17.36 & 4.83E--15 & 0.0446 & (451$\pm$4)E--4   & 0.0460 & 0.0445\\
$^{a}$10:30:25.20 & +33:15:35.6 & 192.819814 &  +59.06935 & 5.11 &   0  &  95 & 20.28 & 18.02 & 3.13E--15 & 0.0162 & (163$\pm$6)E--4   & 0.0175 & 0.0153\\
11:00:32.64       & +22:03:55.4 & 218.197951 &  +64.34454 & 5.18 &   2  & 121 & 20.56 & 17.40 & 4.22E--15 & 0.0185 & (183$\pm$5)E--4  & 0.0190 & 0.0168\\
$^{a}$11:33:39.12 & +13:26:35.9 & 246.056483 &  +67.25447 & 5.35 &   0  & 123 & 20.05 & 17.92 & 3.73E--15 & 0.0425 & (424$\pm$11)E--4  & 0.0450 & 0.0410\\
11:54:07.92       & +27:00:19.1 & 210.208221 &  +77.24893 & 5.03 &   1  & 134 & 20.89 & 17.20 & 4.52E--15 & 0.0257 & (258$\pm$6)E--4  & 0.0273 & 0.0247\\
$^{a}$12:41:44.40 & +21:58:48.7 & 279.229077 &  +84.40000 & 5.10 &   0  & 186 & 19.96 & 17.64 & 6.28E--15 & 0.0293 & (293$\pm$3)E--4   & 0.0298 & 0.0288\\
$^{a}$12:56:12.72 & +17:26:49.2 & 309.674643 &  +80.25619 & 5.01 &   1  & 127 & 20.75 & 17.12 & 5.19E-1-5 & 0.0308 & (306$\pm$5)E--4   & 0.0318 & 0.0295\\
13:34:05.52       & +24:56:42.0 &  22.478206 &  +80.17666 & 5.09 &   1  & 120 & 19.97 & 17.24 & 4.19E--15 & 0.0121 & (120$\pm$5)E--4  & 0.0130 & 0.0110\\
$^{a}$14:02:58.08 & +35:11:28.7 &  65.170753 &  +72.74147 & 5.08 &   3  & 105 & 20.80 & 17.08 & 4.66E--15 & 0.0074 & (79$\pm$5)E--4   & 0.0092 & 0.0074\\
14:22:46.32       & +31:53:33.4 &  51.924015 &  +69.60795 & 5.02 &   3  & 184 & 20.89 & 16.52 & 6.65E--15 & 0.0127 & (129$\pm$6)E--4  & 0.0141 & 0.0120\\
14:55:08.16       & +43:30:10.4 &  74.358140 &  +60.18918 & 5.12 &   3  & 176 & 20.42 & 16.16 & 6.93E--15 & 0.0184 & (184$\pm$4)E--4  & 0.0190 & 0.0176\\
15:31:05.76       & +26:15:17.6 &  40.850494 &  +54.47989 & 5.12 &   9  & 147 & 20.43 & 16.10 & 8.70E--15 & 0.0612 & (59$\pm$2)E--3   & 0.0623 & 0.0554\\
15:53:03.84       & +40:32:32.3 &  64.708720 &  +50.46382 & 5.04 &   11 & 168 & 20.89 & 16.30 & 1.07E--14 & 0.0151 & (155$\pm$9)E--4  & 0.0181 & 0.0144\\
16:30:00.96       & +54:17:46.0 &  83.005830 &  +42.13076 & 5.06 &   21 & 182 & 20.66 & 15.33 & 1.69E--14 & 0.0135 & (138$\pm$7)E--4  & 0.0155 & 0.0127\\
17:08:08.88       & +58:40:45.1 &  87.474735 &  +36.23177 & 5.04 &   20 & 193 & 20.94 & 15.40 & 1.74E--14 & 0.0308 & (305$\pm$16)E--4 & 0.0328 & 0.0274\\
17:25:15.36       & +49:45:58.3 &  76.489792 &  +34.00138 & 5.10 &   26 & 206 & 20.72 & 15.18 & 2.31E--14 & 0.0273 & (269$\pm$17)E--4 & 0.0300 & 0.0235\\
17:57:21.84       & +72:56:51.7 & 103.750419 &  +29.79809 & 5.45 &   49 & 354 & 20.85 & 15.23 & 3.47E--14 & 0.0462 & (454$\pm$6)E--4  & 0.0464 & 0.0439\\
18:28:53.76       & +00:03:40.3 &  30.410269 &  +05.04183 & 5.22 &   2  & 110 & 19.14 & 16.41 & 7.45E--15 & 2.4687 & (249$\pm$16)E--2 & 2.7710 & 2.2336\\
18:58:21.12       & +04:21:07.8 &  37.592645 &  +00.44890 & 5.08 &   29 & 182 & 20.92 & 15.02 & 2.85E--14 & 13.295 & (134$\pm$7)E--1  & 14.409 & 11.864\\
19:31:27.84       & +22:20:38.4 &  57.250887 &  +01.73726 & 5.02 &  168 & 840 & 20.67 & 15.08 & 1.16E--13 & 3.1737 & (310$\pm$11)E--2 & 3.3605 & 2.9270\\
20:15:04.80       & +73:01:26.1 & 106.050051 &  +20.08241 & 5.14 &   55 & 386 & 20.69 & 15.16 & 4.37E--14 & 0.4431 & (45$\pm$3)E--2   & 0.5103 & 0.3994\\
20:43:35.04       & +67:49:17.4 & 102.720707 &  +15.32725 & 5.28 &   22 & 145 & 19.59 & 15.37 & 1.71E--14 & 0.7281 & (71$\pm$4)E--2   & 0.7645 & 0.6330\\
20:59:12.72       & +71:58:57.3 & 107.024821 &  +16.73696 & 5.59 &   29 & 241 & 20.30 & 15.61 & 2.14E--14 & 0.8555 & (86$\pm$5)E--2   & 0.9680 & 0.7814\\
21:39:47.04       & +70:20:33.7 & 108.072504 &  +13.22322 & 5.22 &   31 & 153 & 20.32 & 15.19 & 2.20E--14 & 1.1720 & (111$\pm$6)E--2  & 1.2000 & 0.9856\\
$^{a}$22:00:35.28 & +76:30:26.6 & 113.410573 &  +16.90724 & 5.74 &   10 & 76  & 20.45 & 16.01 & 7.83E--15 & 1.2106 & (118$\pm$2)E--2  & 1.2135 & 1.1261\\
22:21:37.92       & +75:07:08.0 & 113.626695 &  +15.02513 & 5.04 &   14 & 142 & 20.92 & 16.03 & 1.04E--14 & 0.9246 & (89$\pm$3)E--2   & 0.9255 & 0.8383\\
23:06:21.60       & +00:30:48.0 &  76.217222 & --52.55665 & 5.17 &   13 & 127 & 21.05 & 16.43 & 7.79E--15 & 0.0491 & (490$\pm$15)E--4 & 0.0513 & 0.0463\\
23:22:00.72       & +00:38:05.8 &  81.540479 & --54.88397 & 5.02 &   4  & 104 & 20.64 & 16.06 & 5.38E--15 & 0.0403 & (413$\pm$10)E--4 & 0.0441 & 0.0401\\
\noalign{\smallskip} \hline \noalign{\smallskip}

\multicolumn{14}{c}{Southern Hemisphere} \\
00:06:00.55       & --16:15:04.3 &  76.546841 & --74.86283 & 5.06 &  9 & 120 & 20.31 & 16.59 & 6.42E--15 & 0.0256 & (253$\pm$5)E--4  & 0.0260 & 0.0244\\
00:37:40.03       & --29:20:52.4 & 355.928862 & --86.24195 & 5.05 &  3 & 127 & 21.86 & 17.08 & 4.04E--15 & 0.0196 & (198$\pm$15)E--4 & 0.0230 & 0.0178\\
$^{a}$00:59:31.22 & --11:20:16.4 & 130.179575 & --74.09580 & 5.13 &  0 &  75 & 20.37 & 17.57 & 2.91E--15 & 0.0317 & (319$\pm$15)E--4  & 0.0354 & 0.0300\\
01:39:26.57       & --32:50:56.8 & 244.671599 & --78.14690 & 5.07 &  0 &  88 & 21.59 & 18.16 & 1.75E--15 & 0.0250 & (249$\pm$5)E--4  & 0.0259 & 0.0241\\
02:09:49.25       & --00:31:02.3 & 161.438111 & --57.40772 & 5.31 &  0 & 129 & 21.23 & 18.28 & 4.37E--15 & 0.0271 & (276$\pm$8)E--4  & 0.0293 & 0.0265\\
02:33:30.00       & --13:25:28.2 & 188.126797 & --62.51151 & 5.40 &  4 & 115 & 20.50 & 17.39 & 5.51E--15 & 0.0209 & (207$\pm$4)E--4  & 0.0213 & 0.0196\\  
03:08:50.59       & --38:38:24.0 & 243.631348 & --59.16465 & 5.17 &  4 & 123 & 20.86 & 17.06 & 4.17E--15 & 0.0219 & (221$\pm$4)E--4  & 0.0231 & 0.0213\\
03:36:06.19       & --20:58:11.3 & 212.743112 & --52.07320 & 5.06 &  3 & 132 & 19.63 & 16.47 & 8.00E--15 & 0.0273 & (274$\pm$8)E--4  & 0.0286 & 0.0258\\
04:02:24.96       & --38:14:34.1 & 241.007990 & --48.78829 & 5.00 &  9 & 121 & 21.93 & 16.15 & 6.99E--15 & 0.0051 & (49$\pm$5)E--4   & 0.0062 & 0.0040\\
04:20:04.90       & --15:16:00.5 & 210.263825 & --40.297736& 5.12 &  9 & 108 & 19.69 & 16.34 & 8.46E--15 & 0.0573 & (56$\pm$3)E--3   & 0.0607 & 0.0502\\
05:01:40.01       & --48:54:27.7 & 255.330888 & --37.82428 & 5.04 &  16& 159 & 21.92 & 15.71 & 1.20E--14 & 0.0117 & (115$\pm$4)E--4  & 0.0123 & 0.0106\\
$^{a}$05:41:55.18 & --08:27:23.4 & 212.698747 & --19.26914 & 5.03 &  0 &  15 & 19.69 & 18.27 & 5.90E--16 & 3.5394 & (351$\pm$9)E--2   & 3.7273 & 3.2823\\
05:57:53.33       & --13:37:16.7 & 219.388114 & --17.91770 & 5.14 &  30& 170 & 20.91 & 15.22 & 2.63E--14 & 0.5327 & (55$\pm$5)E--2   & 0.6337 & 0.4742\\
09:05:29.28       & --06:00:11.6 & 235.431491 &  +26.16744 & 5.10 &  29& 161 & 19.40 & 15.01 & 2.86E--14 & 0.0281 & (282$\pm$9)E--4  & 0.0299 & 0.0267\\
09:31:44.64       & --01:45:46.9 & 235.687810 &  +33.96480 & 5.14 &  25& 210 & 20.97 & 15.57 & 1.96E--14 & 0.0336 & (336$\pm$9)E--4  & 0.0353 & 0.0315\\
10:10:14.16       & --02:15:46.8 & 243.559329 &  +41.31715 & 5.02 &  24& 238 & 20.47 & 16.03 & 1.70E--14 & 0.0408 & (407$\pm$10)E--4 & 0.0428 & 0.0391\\
10:26:21.12       & --00:07:38.5 & 244.926543 &  +45.76736 & 5.07 &  16& 96  & 19.97 & 16.12 & 9.86E--15 & 0.0541 & (552$\pm$19)E--4 & 0.0591 & 0.0527\\
11:05:33.84       & --77:46:21.7 & 297.294235 & --16.07566 & 5.18 &  3 & 99  & 20.98 & 16.37 & 4.39E--15 & 1.9375 & (191$\pm$16)E--2 & 2.1494 & 1.5896\\
11:23:08.16       & --01:42:10.3 & 263.002375 &  +54.17694 & 5.18 &  6 & 113 & 19.87 & 16.40 & 5.95E--15 & 0.0659 & (66$\pm$3)E--3   & 0.0713 & 0.0611\\
11:58:47.28       & --03:42:52.4 & 278.523051 &  +56.64133 & 5.01 &  5 & 132 & 19.82 & 16.63 & 6.91E--15 & 0.0294 & (296$\pm$12)E--4 & 0.0326 & 0.0277\\
12:19:55.20       & --00:09:51.1 & 286.137714 &  +61.67555 & 5.11 &  7 & 107 & 20.47 & 16.48 & 5.79E--15 & 0.0231 & (234$\pm$4)E--4  & 0.0244 & 0.0229\\
12:55:39.12       & --00:47:28.7 & 305.180788 &  +62.06203 & 5.05 &  12& 94  & 20.98 & 16.43 & 6.25E--15 & 0.0175 & (182$\pm$9)E--4  & 0.0203 & 0.0168\\
13:38:06.24       & --10:44:17.5 & 321.123566 &  +50.47792 & 5.07 &  7 & 156 & 19.82 & 16.12 & 9.47E--15 & 0.0810 & (80$\pm$4)E--3   & 0.0856 & 0.0709\\
14:00:15.84       & --03:33:17.4 & 334.006917 &  +55.10961 & 5.09 &  13& 178 & 20.18 & 16.06 & 1.20E--14 & 0.0462 & (47$\pm$2)E--3   & 0.0520 & 0.0435\\
14:37:44.40       & --05:14:50.0 & 345.355145 &  +48.67014 & 5.18 &  19& 160 & 21.72 & 16.01 & 1.18E--14 & 0.0719 & (724$\pm$3)E--4  & 0.0729 & 0.0718\\
15:04:41.52       & --03:47:04.5 & 354.237024 &  +45.39982 & 5.06 &  22& 152 & 20.51 & 15.59 & 1.81E--14 & 0.1748 & (173$\pm$7)E--3  & 0.1839 & 0.1608\\
15:42:55.44       & --34:06:39.6 & 338.915415 &  +16.51832 & 5.02 &  8 & 77  & 19.05 & 16.36 & 7.81E--15 & 0.8542 & (861$\pm$15)E--3 & 0.9031 & 0.8371\\
$^{a}$15:53:31.92 & --04:42:22.8 &   4.103063 &  +35.73414 & 5.54 &  0 &  41 & 20.95 & 18.08 & 1.26E--15 & 0.9739 & (93$\pm$6)E--2    & 1.0348 & 0.8230\\
16:27:41.76       & --24:45:20.9 & 352.989963 &  +16.46125 & 5.08 &  0 &  22 & 20.80 & 18.89 & 5.13E--16 & 2.3699 & (23$\pm$5)E--1   & 3.4392 & 1.6456\\
16:50:49.68       & --15:22:04.8 &   4.219387 &  +18.07220 & 5.07 &  0 &  72 & 19.99 & 18.19 & 2.70E--15 & 1.7061 & (169$\pm$11)E--2 & 1.8476 & 1.5248\\
17:34:12.72       & --25:23:06.0 &   1.665072 &  +04.03459 & 5.48 &  21& 282 & 19.73 & 15.43 & 2.73E--14 & 2.6267 & (261$\pm$5)E--2  & 2.6868 & 2.4711\\
18:04:55.92       & --24:17:11.8 &   6.191747 & --01.37869 & 5.29 &  0 & 2   & 17.73 & 17.64 & 3.11E--16 & 7.8239 & (8$\pm$2)E--0    & 12.452 & 5.7035\\
$^{a}$18:28:03.12 & --03:49:55.9 &  26.846166 &  +03.43861 & 5.77 &  0 & 30  & 20.16 & 18.08 & 1.44E--15 & 10.819 & (10$\pm$5)E--0    & 23.146 & 3.8047\\
19:04:04.08       & --37:17:30.1 & 359.760820 & --18.377007& 5.01 &  11& 144 & 20.83 & 16.10 & 9.96E--15 & 0.9750 & (96$\pm$5)E--2   & 1.0277 & 0.8332\\
19:32:23.76       & --78:26:14.3 & 315.912763 & --28.58364 & 5.57 &  38& 191 & 20.08 & 15.07 & 2.69E--14 & 0.4874 & (48$\pm$4)E--2   & 0.5469 & 0.4208\\
19:52:11.04       & --81:06:33.8 & 312.761609 & --29.113235& 5.51 &  38& 189 & 20.87 & 15.14 & 2.58E--14 & 0.7702 & (73$\pm$8)E--2   & 0.8514 & 0.5749\\
21:05:41.76       & --27:46:21.0 &  18.361528 & --40.37909 & 5.24 &  43& 302 & 20.23 & 15.00 & 3.33E--14 & 0.1084 & (107$\pm$2)E--3  & 0.1113 & 0.1026\\
21:22:02.16       & --07:17:39.3 &  44.676556 & --36.66431 & 5.01 &  15& 166 & 21.00 & 15.63 & 1.32E--14 & 0.2486 & (241$\pm$6)E--3  & 0.2497 & 0.2243\\
22:07:12.72       & --19:14:43.4 &  35.818926 & --51.61918 & 5.02 &  12& 145 & 21.13 & 16.38 & 1.01E--14 & 0.0260 & (258$\pm$10)E--4 & 0.0286 & 0.0244\\
$^{a}$22:41:36.24 & --13:28:34.7 &  50.650654 & --56.77675 & 5.31 &  16&  151 & 21.00 & 16.07 & 1.01E--14 & 0.0495 & (497$\pm$17)E--4 & 0.0527 & 0.0464 \\  
22:58:27.84       & --47:34:36.8 & 342.524449 & --59.94059 & 5.25 &  8 & 155 & 21.88 & 16.33 & 6.34E--15 & 0.0097 & (102$\pm$12)E--4 & 0.0130 & 0.0086\\
$^{a}$23:40:51.12 & --03:53:58.6 &  83.895230 & --61.30129 & 5.25 &  10&  90  & 20.65 & 16.62 & 6.10E--15 & 0.0373 & (374$\pm$12)E--4 & 0.0399 & 0.0359\\

\noalign{\smallskip}\hline\noalign{\smallskip}

\multicolumn{14}{l}{{\bf Note:} $^{a}$ DBF validated and regularly used at the GTC.} \\

\end{longtable}

\twocolumn
\end{landscape}
\normalsize


\label{lastpage}

\end{document}